\documentclass[a4paper,12pt]{article}
\begin{document}

{\noindent \sf \large Coulomb transition matrix at negative energy and \\
integer values of interaction parameter } \\[.2in]
{\sf V. F Kharchenko}\\[.1in] 
{\footnotesize Bogolyubov Institute for Theoretical Physics,  
National Academy of Sciences \\of Ukraine, UA - 03143, Kyiv, Ukraine \\ [.1in]
E-mail: vkharchenko@bitp.kiev.ua} \\[.1in]
\noindent \small{ {\sf Abstract }  \\ 
{\small With the use of the stereographic projection of momentum space into the 
four-dimensional sphere of unit radius. the possibility of the analytical solution of 
the three-dimensional two-body Lippmann-Schwinger equation with the Coulomb 
interaction at negative energy has been studied. Simple analytical expressions for 
the three-dimensional Coulomb transition matrix in the case of the repulsive  
Coulomb interaction and positive integer values of the Coulomb parameter have 
been obtained. The worked out method has been also applied for  the generalized 
three-dimensional Coulomb transition matrix in the case of the attractive Coulomb 
interaction and negative integer values of the Coulomb parameter. } \\ [.05in]

\noindent {\sf 1. Introduction} \\ 

\noindent It is known that the Coulomb transition matrix for two charged particles, 
which interact with each other, $t^C(E)$, as well as connected with it the Coulomb Green 
function 
\begin{equation}
g^C(E)=g^0(E)+g^0(E)t^C(E)g^0(E),
\end{equation}  
where $g^0(E)$ is the free Green's function ($E$ is the energy of the system), contain 
all the information about the two-body system. The two-body Coulomb transition matrix 
off the energy shell is an important quantity in both the atomic and nuclear physics. 
It immediately appears in the integral kernels of the Faddeev equations [1] describing 
the three-particle system with two or all three particles of which are charged. 

In the literature, a number of the different representations of the Coulomb Green's 
function (see the review [2]). The representation, which allows explicitly the known 
symmetry of the Coulomb system in the four-dimensional Euclidean space, firstly revealed 
by Fock [3], is particularly perspective. With the help of the stereographic projection 
of the momentum space onto the four-dimensional sphere of the unit radius, 
a single-parameter integral representation of the three-dimensional Coulomb Green's 
function has been derived in the papers by Bratsev and Trifonov [4] and Schwinger [5].
With the use of the Fock symmetry, the general expressions for the three-dimensional 
Coulomb transition matrix with explicitly separated singularities in the transfer 
momentum and the energy have been derived in the papers [6] (in the case of the negative 
energy) and [7] (in the case of zero and positive energies).

The possibility of the derivation of a simple analytical expression for the partial 
Coulomb transition matrix, which obey the Lippman-Schwinger equation with the energy of 
the ground bound state of the two-body complex, has been firstly demonstrated in the 
papers [8] (in the case of the attractive interaction with the Coulomb parameter 
$\gamma=-1$) and [9] (in the case of the repulsive interaction with the parameter 
$\gamma=1$). The analytical expressions for the two-particle partial Coulomb transition 
matrices  at the energy of the first-excited state ($\gamma=2$ and $\gamma=-2$) have been 
obtained in [10].

In this paper, leaning upon the formalisms developed in [5] and [6], we firstly find the 
possibility of the analytical solution of the Lippman-Schwinger equation for the 
three-dimensional Coulomb transition matrices in the case of the repulsive interaction 
at integer values of the Coulomb parameter $\gamma$.

The worked out method has also been applied for the generalized Coulomb transition matrix 
in the case of the attractive Coulomb interaction and integer values of the Coulomb 
parameter.

In the Section 2,  the basic formulas for the three-dimensional Coulomb transition 
matrices obtained with the use of the Fock method of the stereographic projection are 
presented. The Section 3 is devoted to the derivation of the analytical expressions for 
the three-dimensional Coulomb transition matrix in the case of the repulsive Coulomb 
interaction with positive integer Coulomb parameter. In the Section 4, analytical 
expressions for the generalized Coulomb transition matrix with negative integer values 
of the Coulomb parameter. he ection 5 is devoted to the discussion and conclusions. \\ 

\noindent {\sf 2. Fock method}\\ 

In the momentum representation the three-dimensional Coulomb transition matrix 
$<{\bf k}|t^C(E)|{\bf k}^{\prime}>$ satisfies the Lippmann-Schwinger equation
\begin{equation}
<{\bf k}|t^C(E)|{\bf k}^{\prime}>=\langle {\bf k} \mid v^C \mid {\bf k}^{\prime}\rangle + 
\int \frac{d{\bf k}^{\prime\prime}}{(2\pi)^3}\langle {\bf k} \mid v^C \mid {\bf k}^{\prime\prime}\rangle
\frac{1}{E-\frac{k^{\prime\prime 2}}{2\mu}} <{\bf k}^{\prime\prime}|t^C(E)|{\bf k}^{\prime}>\;\;,   
\end{equation}  
where $E$, ${\bf k}$ and ${\bf k}{\prime}$ are the energy and the momenta of the relative 
motion of two particles, $\mu$ is the reduced mass of the particles,
$\langle {\bf k} \mid v^C \mid {\bf k}^{\prime}\rangle$ is the matrix of the Coulomb 
interaction of the particles 1 and 2,
\begin{equation}
<{\bf k}|v^C|{\bf k}^{\prime}>=\frac{4\pi q_1 q_2}{\mid {\bf k}-{\bf k}^{\prime}\mid ^2}\;,
\end{equation} 
$q_i$ is the charge of the particle. In this paper we restrict our consideration of the 
two-particle system with the negative energy
\begin{equation}
E=-\frac{\hbar^2 \kappa^2}{2\mu}\;,
\end{equation} 
$\hbar$ is the reduced Planck constant.

Using the Fock method of the stereographic projection of the momentum space onto the 
four-dimension sphere with the unit radius [3], the solution of the integral equation 
for the three-dimensional Coulomb transition matrix at the negative energy ($E<0$)(2) 
can be written in the form of the sum [4]
\begin{equation}
<{\bf k}|t^C(E)|{\bf k}^{\prime}>=\frac{2\pi q_1 q_2 \eta}{k k'}
\left[ \frac{1}{\sin^2 \frac{\omega}{2}} - \frac{4\gamma}{\sin \omega} \sum_{n=1}^{\infty} 
\frac{\sin n\omega}{n + \gamma}\right]\;,
\end{equation}
the integral representation [4,5]
\begin{equation}
<{\bf k}|t^C(E)|{\bf k}^{\prime}>=\frac{2\pi q_1 q_2 \eta}{k k'}
\left[ \frac{1}{\sin^2 \frac{\omega}{2}} - 4\gamma \int_{0}^{1} d\rho 
\frac{\rho^{\gamma}}{\rho^2 -2 \cos \omega \cdot \rho + 1} \right]\;,
\end{equation}
and with explicitly separated singularities in the transfer momentum and the energy [6]
\begin{displaymath}
<{\bf k}|t^C(E)|{\bf k}^{\prime}>=\frac{2\pi q_1 q_2 \eta}{k k'}
\left\{ \frac{1}{\sin^2 \frac{\omega}{2}} - \frac{2}{\sin \omega} \left[ \pi \gamma 
\cos \gamma \omega + \gamma \sin 2\gamma\omega\cdot \ln\left( \sin\frac{\omega}{2} \right) \right.\right.
\end{displaymath}
\begin{equation}
\left.\left. - 2\pi \gamma c(\gamma) \cot \gamma\pi\cdot\sin \gamma\omega - \gamma \cos\gamma\omega \cdot
x_{\gamma}(\omega) - 2\gamma^2 \sin\gamma\omega \cdot y_{\gamma}(\omega)\right]\right\}\;,
\end{equation}
where
\begin{equation}
\eta = \frac{2\kappa^2 kk'}{\left(k^2 + \kappa^2 \right)\left(k'^2 + \kappa^2 \right)}\;,
\end{equation}
$\gamma$ is the known dimensionless Coulomb parameter
\begin{equation}
\gamma = \frac{\mu g_1 q_2}{\hbar^2 \kappa}
\end{equation}
and the variable quantity $\omega$ denotes the angle between two vectors in the 
four-dimensional space introduced by Fock [3], 
\begin{equation}
\sin^2\frac{\omega}{2} = \frac{{\kappa^2}\mid {\bf k} - {\bf k}^\prime \mid ^2}
{(k^2 + \kappa^2)(k^{\prime 2} + \kappa^2)}\;\; , \;\; 0\leq \omega < \pi\;\;.
\end{equation} 
The functions  $x_{\gamma}(\omega)$, $y_{\gamma}(\omega)$ and $c(\gamma)$ in 
(7) are given by the expressions 
\begin{displaymath}
x_{\gamma}(\omega)= \int_{0}^{\omega} d\varphi \;\sin \gamma\varphi\;
\cot \frac{\varphi}{2} \;,\;\;\; y_{\gamma}(\omega)= \int_{\omega}^{\pi} 
d\varphi \;\sin \gamma\varphi\;\ln\left(\sin \frac{\varphi}{2}\right)\;,
\end{displaymath}
\begin{equation}
c(\gamma)= \frac{1}{2} \left[ 1 - \frac{1}{\pi} x_{\gamma}(\pi)\right]\;.
\end{equation} 
The energy $E_{\gamma}$, which corresponds to the given Coulomb parameter 
$\gamma$ (9),is equal to
\begin{equation}
E_{\gamma} = -\frac{\hbar^2 \kappa^2}{2\mu}= - \frac{\mu (q_1 q_2)^2}
{2 \hbar^2 \gamma^2}\;. 
\end{equation}

If the Coulomb potential is attractive (the particles have the charges of opposite 
signs, $g_1 q_2<0$), the Coulomb parameter $\gamma$ takes the negative value, 
$\gamma<0$. To the spectrum of the bound states of the two-particle system with 
the energies
\begin{equation}
E_n = - \frac{\mu (q_1 q_2)^2}{2 \hbar^2 n^2}\;, \qquad  n=1, 2, \cdots \;,
\end{equation}
correspond the negative integer values $\gamma$ ($\gamma 
= -n,\; n=1,2,3,...$) and, according to (4), the values
\begin{equation}
\kappa_n = \frac{\mu \mid q_1 q_2 \mid }
{\hbar^2 n}\;.
\end{equation}

The values of the Coulomb parameter (9) at $\kappa=\kappa_n$ is integer
\begin{equation}
\gamma_n = \frac{q_1 q_2}{\mid q_1 q_2 \mid}n
\end{equation}
- positive for the repulsive Coulomb interaction ($q_1 q_2>0$, $\gamma_n = n$) 
and negative for the attractive interaction ($q_1 q_2<0$, $\gamma_n = - n$).

In view of that in the expression (11) for $c(\gamma)$ the quantity $x_{\gamma}(\pi)$
with the integer values of $\gamma$ takes the values
\begin{equation}
x_n(\pi) = \pi\;,\qquad   x_{-n}(\pi) = - \pi\;,
\end{equation}
we find that in the expression (7)
\begin{equation}
c(n) = 0\;,\qquad   c(-n) = 1\;.
\end{equation}

Hence, in the case of the attractive Coulomb interaction ($\gamma<0$), the third term 
in the square brackets of the expression (7) through the presence of the function 
$\cot \gamma\pi$ contains the singularities at $\gamma = \gamma_n$, which correspond 
to the energies of bound states $E=E_n$.

If the Coulomb interaction is repulsive ($\gamma>0$), the expression for the Coulomb 
transition matrix (7) does not contain singularities at the energies of bound states. 
Evaluating in this case in the third term of the expression in (7) the indefinite form 
of the type $\frac{0}{0}$ at $\gamma=\gamma_n$ according to the l'Hospital rule, we find 
\begin{equation}
\left[ \frac{2\pi \gamma c(\gamma)}{\tan \gamma\pi}\right]_{\gamma \rightarrow n} 
= 2 n c^\prime (n)\equiv \rho_n\;,
\end{equation}
where
\begin{equation}
\rho_n = (-1)^n - 2 n \ln 2 - 2 n \sum_{m=1}^{n} \frac{(-1)^m}{m} \;.
\end{equation}
\\ 

\noindent {\sf 3. Three-dimensional Coulomb transition matrix in the case of
positive integer value of the Coulomb parameter}\\ 

Using the formula (7), consider the expression for the three-dimensional Coulomb 
transition matrix in the case of the particles with the charges of the same sign 
at the energy $E=E_n$ (13). In this case the parameter $\gamma$ takes the positive 
integer values (15).

When $\gamma=1$ we have in accordance with (11)
\begin{equation}
x_1(\omega) =  \omega + \sin\omega \;, \;\;\;
y_1(\omega) = - \cos^2 \frac{\omega}{2}\;- \;
\sin^2 \frac{\omega}{2}\; \ln\sin^2\left(  \frac{\omega}{2} \right)\;,\;\\
\end{equation}
\begin{displaymath}
c(1)=0\;,\;\;\; \rho_1=1-2\ln2\;.
\end{displaymath} 
The three-dimensional transition matrix then takes the simple analytical form
\begin{equation}
\langle {\bf k} \mid t(E) \mid {\bf k}^{\prime}\rangle=\frac{2\pi q_1 q_2 \eta}{k k'}
\left\{ \frac{1}{\sin^2 \frac{\omega}{2}} - 2(\pi - \omega)\cot\omega - 
2\ln \sin^2\frac{\omega}{2}) -4\ln2 \right\} \;.
\end{equation} 

In an analogous way, using (11) with $\gamma  = 2$ we find
\begin{equation}
x_2(\omega) =  \omega + 2\sin\omega + \frac{1}{2} \sin 2\omega \;,\;\;\;
y_2(\omega) = -\cos^4 \frac{\omega}{2}\;-\;
\frac{1}{2}\sin^2 \frac{\omega}{2}\; \ln\sin^2\left(  \frac{\omega}{2} \right)\;,\\
\end{equation} 
\begin{displaymath}
c(2)=0\;,\;\;\; \rho_2=3-4\ln2
\end{displaymath}
and the expression for the three-dimensional Coulomb transition matrix at $E=E_2$ 
in the form
\begin{equation}
\langle {\bf k} \mid t(E) \mid {\bf k}^{\prime}\rangle=\frac{2\pi q_1 q_2 \eta}{k k'}
\left\{ \frac{1}{\sin^2 \frac{\omega}{2}} - 4(\pi - \omega)\frac{\cos 2\omega}
{\sin \omega} - 8\cos \omega\cdot \ln \sin^2\frac{\omega}{2}) - 16 \ln2\cdot \cos \omega - 
8 \right\}.
\end{equation} 

At   $\gamma  = 3$ we have
\begin{displaymath}
x_3(\omega) =  \omega + 2\sin\omega + \frac{1}{3} \sin 3\omega \;,\\  
\end{displaymath}
\begin{equation}
y_3(\omega) = - \left(\frac{16}{3} \sin^6\frac{\omega}{2} -8\sin^4\frac{\omega}{2} + 
3 \sin^2\frac{\omega}{2}\right)\ln \sin^2\frac{\omega}{2} 
\end{equation} 
\begin{displaymath}
+ \frac{16}{9} \sin^6\frac{\omega}{2} - 4\sin^4\frac{\omega}{2} + 
3\sin^2\frac{\omega}{2} - \frac{7}{9}, \;\;\;
c(3)=0\;,\;\; \rho_3=4-6\ln2,
\end{displaymath}
and the corresponding Coulomb t-matrix at $E=E_3$ is equal to 
\begin{equation}
\langle {\bf k} \mid t(E) \mid {\bf k}^{\prime}\rangle=\frac{2\pi q_1 q_2 \eta}{k k'}
\left\{ \frac{1}{\sin^2 \frac{\omega}{2}} - 6(\pi - \omega)\frac{\cos 3\omega}
{\sin \omega} - 6\left( 2\cos 2\omega + 1 \right) \ln \sin^2\frac{\omega}{2}) \right.
\end{equation} 
\begin{displaymath}
\left.- 24 \ln2\cdot \cos 2\omega - 24\cos \omega -12\ln 2 - 6 \right\}.
\end{displaymath}\\ 

\noindent {\sf 4. Generalized Coulomb transition matrix in the case of
negative integer Coulomb parameter $\gamma = -1$}\\ 

In the case of the system with the attractive Coulomb interaction, of special 
interest is known as the generalized Coulomb Green function [4,11] at the energy 
of the bound ground state $E=E_1=-b_1$,  $\tilde{g}^C(E_1)$. It related with the
generalized Coulomb transition matrix  $\tilde{t}^C(E_1)$ by the relation
\begin{equation}
\tilde{g}^C(E_1)=g^0(E_1)+g^0(E_1)\tilde{t}^C(E)g^0(E),
\end{equation}  

The expression for $\tilde{t}^C(E_1)$ follows from the expression (5) for $t^C(E_1)$ excluding 
the singular term at $n=1$, which describes the contribution from ground state,
\begin{equation}
<{\bf k}|\tilde{t}^C(E)|{\bf k}^{\prime}>=\frac{2\pi q_1 q_2 \eta}{k k'}
\left[ \frac{1}{\sin^2 \frac{\omega}{2}} - \frac{4\gamma}{\sin \omega} \sum_{n=2}^{\infty} 
\frac{\sin n\omega}{n + \gamma}\right]\;,
\end{equation}
or, in view of Eq. (6)
\begin{equation}
<{\bf k}|{\tilde{t}}^C(E)|{\bf k}^{\prime}>=\frac{2\pi q_1 q_2 \eta}{k k'}
\left[ \frac{1}{\sin^2 \frac{\omega}{2}} - 4\gamma \left( \int_{0}^{1} d\rho 
\frac{\rho^{\gamma}}{{\rho^2}-2\cos \omega \cdot \rho + 1} - \frac{1}{1+\gamma}\right) \right]\;.
\end{equation}

Writing $(1+\gamma)^{-1}$ inthe form of the definite integral
\begin{equation}
\frac{1}{1+\gamma} = \int_{0}^{1} d\rho \rho^{\gamma}\;,
\end{equation}
we obtain the following expression for the generalized transition matrix
\begin{equation}
<{\bf k}|\tilde{t}^C(E)|{\bf k}^{\prime}>=\frac{2\pi q_1 q_2 \eta}{k k'}
\left[ \frac{1}{\sin^2 \frac{\omega}{2}} + 4\gamma \int_{0}^{1} d\rho \rho^{\gamma + 1}
\frac{\rho -2 \cos \omega}{\rho^2 - 2 \cos \omega\cdot \rho + 1}\right]\;.
\end{equation}

At the energy of the ground bound state $E=-b_1$ we have $\gamma_1=-1$ and the integral 
in (30) is equal to
\begin{equation}
\int_{0}^{1} d\rho \frac{\rho -2 \cos \omega}{\rho^2 - 2 \cos \omega \rho + 1}=
\left( \frac{\omega}{2}-\frac{\pi}{2}\right) cot \omega  + \ln |2 \sin \frac{\omega}{2}|\;.
\end{equation}

As a result, for the generalized Coulomb transition matrix at the energy of the ground 
bound state (and the corresponding values $\kappa = \kappa_1, \gamma = \gamma_1, \omega = 
\omega_1$) we find
\begin{equation}
<{\bf k}|{\tilde{t}}^C(-b_1)|{\bf k}^{\prime}>=\frac{2\pi q_1 q_2 \eta}{k k'}
\left[ \frac{1}{\sin^2 \frac{\omega}{2}} + 2 \left( \pi - \omega\right) cot \omega - 
4 \ln \left|2 \sin \frac{\omega}{2} \right|\right]\;.
\end{equation}
\\ 

\noindent {\sf 5. Discussion and conclusions.}\\ 

The study of properties of the two-particle Coulomb transition matrix is the topical problem, 
which arises in both the atomic and nuclear physics in connection with formulation and 
solution of the equations describing few-body systems which contain the charged particles.
A knowledge of the two-particle Coulomb transition matrix is necessary in particular to 
determine the electric $2^{\lambda}$-pole polarizabilities of the two-body atomic and nuclear 
systems [12, 13].

In specific cases, the analytical solution of the Lippmann-Schwinger equation for the partial 
two-body Coulomb transition matrices has been realizeed by us previously using the symmetry of 
the Coulomb systems in the Fock four-dimensional space - in [8,9] (the system with unlike 
charges) and in [10] (the system with like charges).

In the present paper, it has been firstly shown the possibility of the analytical solution of 
the equationn for the three-dimensional Coulomb transition matrix in the case of the repulsive 
interaction with the positive integer values of the Coulomb parameter.

The obtained analytical expressions for the three-dimensional Coulomb transition matrix (21), 
(23) and (25) are in agreement with the corresponding formulae for the partial Coulomb 
transition matrices
\begin{equation}
t_l^C(k,k^{\prime};E_n) = \frac{1}{4\eta} \int_{\omega_0}^{\omega_{\pi}} d\omega\; 
\sin \omega \; P_l \left( \frac{2\xi-1 + \cos \omega}{2\eta} \right)
\langle {\bf k} \mid t^ (E) \mid {\bf k}^{\prime}\rangle  \;,
\end{equation}
where $P_l(x)$ is the Legendre polynom,
\begin{equation}
\xi = \frac{\kappa^2 (k^2 + {k^{\prime}}^2)}{(k^2 + \kappa^2)({k^{\prime}}^2 + 
\kappa^2)}\;,\quad \omega_0 = 2 \arcsin \sqrt{\xi - \eta}\;, \qquad \; 
\omega_{\pi} = 2 \arcsin \sqrt{\xi + \eta}\;,
\end{equation} 
obtained by us earlier [10] in the case of the repulsive Coulomb interaction.

In the case of the attractive interaction, it has been performed analytical solution 
of the equation for the generalized Coulomb transition matrix with the negative integer value 
of the Coulomb parameter $\gamma=-1$.

Note, that that the result (31) obtained by us for the generalized Coulomb transition matrix 
$<{\bf k}|{\tilde{t}}^C(-b_1)|{\bf k}^{\prime}>$ substantially differs from the result off 
Bratsev and Trifonov [4] as a consequence of the error when integrating in [4]. \\

\noindent {\sf Acknowledgment}\\ 

The present work was partially supported by the National Academy of Sciences 
of Ukraine (project No. 0117U00237) and by the Program of Fundamental Research 
of the Department of Physics and Astronomy of NASU (project No. 0117U00240). 
\\ [.2in] 

\noindent {\footnotesize {\sf References} 
\vspace*{.1in}
\begin{itemize} 
\setlength{\baselineskip}{.1in} 
\item[{\tt [1]}] Faddeev L. D. 1961 Sov. Phys. JETP 12 1014-1019. 
\item[{\tt [2]}] Chen J. C. Y. and Chen A. C. 1972 Advances in Atomic and Molecular Physics 
                  ed D. B. Bates and I. Estermann vol 8 (N Y - London: Academic Press)pp 71-129.
\item[{\tt [3]}] Fock V. A. 1935 Z. Phys. 98 145-154. 
\item[{\tt [4]}] Bratsev V. F. and Trifonov  E. D. 1962 Vest. Leningrad. Gos. Univ. 16 36-39. 
\item[{\tt [5]}] Schwinger J. 1964 J. Math. Phys. 5 1606-1608.
\item[{\tt [6]}] Shadchin S. A. and Kharchenko V. F. 1983 J. Phys. B: At. Mol.Phys. 16 1319-1322. 
\item[{\tt [7]}] Storozhenko S. A. and Shadchin S. A. 1988 Teor. Mat. Fiz. 76 339-349. 
\item[{\tt [8]}] Kharchenko V. F. 2016 Ann. Phys. NY 374 16-26.
\item[{\tt [9]}] Kharchenko V. F. 2018 Canadian J. Phys. 96 933-937.
\item[{\tt [10]}] Kharchenko V. F. 2017 Ukrainian J. Phys. 62 263-70.
\item[{\tt [11]}] Sherstyuk A. I. 1971 Teor. Mat. Phys. 7 342-347.
\item[{\tt [12]}] Kharchenko V. F. 2013 J. Mod. Phys. 4 99-107.

\end{itemize} 

\end{document}